\documentclass[ reprint,
superscriptaddress,
nobibnotes,
 amsmath,amssymb,
 aps,
prx,onecolumn
]{revtex4-2}

\usepackage{graphicx}
\usepackage{dcolumn}
\usepackage{bm}
\usepackage{graphicx,epsfig,epstopdf}
\usepackage{float}
\usepackage{subcaption}
\usepackage{amsmath} 
\usepackage{bm}
\usepackage{color}
\usepackage{caption}
\usepackage{amssymb} 
\DeclareMathAlphabet{\mathcal}{OMS}{cmsy}{m}{n}
\newcommand{\be}{\begin{eqnarray}}
\newcommand{\ee}{\end{eqnarray}}
\newcommand{\bc}{\begin{center}}
\newcommand{\ec}{\end{center}}

\newcommand{\nn}{\nonumber}

\newcommand{\ket}[1]{ |#1\rangle}
\newcommand{\bra}[1]{\langle #1|}

\newcommand{\comm}[2]{[\,#1,#2\,]}

\usepackage{centernot}
\usepackage{cancel}

\begin{document}

\preprint{APS/123-QED}
\title{Coherent nonlinear optical probe for cavity-dressed vibrational mode mixing: multidimensional double-quantum coherence and photon-echo spectroscopy}

\author{Arunangshu Debnath}
\email{arunangshu.debnath@desy.de}
\affiliation{Theory Division, Center for Free-Electron Laser Science CFEL, Deutsches Elektronen-Synchrotron DESY, Notkestrasse 85, 22607 Hamburg, Germany}

\begin{abstract}
Cavity dressing of molecular vibrational dynamics expands the role of characteristic vibrations as spectroscopic markers of underlying ultrafast dynamics. Interacting vibrational modes exhibit a pronounced excited state delocalization due to the interaction with the cavity mode, which is reflected in the ultrafast dynamics. 
We characterize the ultrafast dynamics of these cavity-dressed characteristic vibrations in the presence of dissipation. 
Specifically, we present two complementary three-pulse coherent multidimensional spectroscopic techniques capable of monitoring one- and two-quantum cavity-dressed vibrational excitations. Dissipative properties, such as transport and dephasing, are described using a microscopic theory that includes low- and high-energy phonon modes.
Simulations were performed with finite laser pulses. The cavity coupling strengths fall within a range similar to vibrational mode couplings, hinting towards a possibility of control of intermolecular vibrational energy redistribution. The framework is extendable to a broad range of cavity-controlled nonlinear spectroscopies of dissipative molecular systems.
\end{abstract}

\maketitle

\section{Introduction}\label{sec:intro}
Vibrational modes with distinct spectral signatures in the IR regime routinely serve as spectroscopic markers. They allow the monitoring of underlying ultrafast molecular dynamics. However, when two or more of these characteristic modes interact with each other and with auxiliary vibrations, the vibrational response encodes both collective and dissipative spectral features. Consequently, assigning spectroscopic modes and disentangling dynamical features from the collective response becomes challenging.
In general, time-resolved nonlinear spectroscopies can monitor ultrafast vibrational dynamics in the presence of dissipation \cite{khalil2003coherent, demirdoven2002correlated, baiz2020vibrational}. These techniques, collectively referred to as multidimensional coherent spectroscopy (MDCS), utilize multiple, well-controlled, coherent laser–matter interactions to probe the spectral and temporal features of vibrational dynamics \cite{hamm2011concepts, li2023optical, wright2003quantitative, baiz2020vibrational}.
Previous studies have demonstrated the capabilities of multidimensional spectroscopies in exploring solvation dynamics \cite{tanimura1993two, ishizaki2007dynamics}, chemical transformation \cite{mukamel1990femtosecond}, 
and the conformational changes of proteins and peptides \cite{ghosh2017watching}. \\
In recent years, the vibrational modes have been modulated via another tool, the localized photonic modes, broadly referred in the literature as cavity modes \cite{brawley2025vibrational, li2021cavity, dunkelberger2022vibration, li2022polariton, schafer2022shining, xiang2024molecular, xiong2023molecular, simpkins2023control, mondal2025polariton, schnappinger2023ab, borges2025selective}.
Early spectroscopic studies of cavity-dressed ultrafast vibrational dynamics sought to focus on the strong-coupling regime, characterized by laser detuning-dependence of vibrational mode splitting. The magnitude of these splittings helps estimate cavity coupling strengths, which often fall within a range similar to the vibrational energy gaps. Corroborative theoretical developments usually adopt models inspired by atomic physics counterparts, specifically the Tavis–Cummings model, in which many non-interacting two-level systems couple uniformly to a cavity mode. 
Molecular cavity quantum electrodynamics operates in a regime that is somewhat distinct. Atomic excited states are typically spatially confined. The ratio between the radiative emission of the atoms into the cavity mode and their decay into other channels governs the coupling strength. In contrast, vibrational excited states are spatially delocalized, and the ultrafast dynamics is dominated by dephasing interactions. The cavity coupling and the temporal-spectral properties of the laser source jointly determine the nature of the vibrational excitations that can be probed.
Furthermore, multimode vibrations within a single molecular system inherently host collective excitations. In such cases, a cavity mode can either modulate intramolecular collective excitations or couple intermolecular excitations across multiple molecules. This study focuses on the former case and assumes a dispersive, intermediate-coupling regime. This regime is particularly suitable for spectroscopy because it enables cavity modulation while ensuring that emission into the cavity mode remains low enough to allow for the generation of strong nonlinear spectroscopic signals.\\
Selected previous works focusing on first-principles simulations also hints toward a more general picture: in describing the cavity modulation of collective vibrational modes, the multi-level nature of the vibrations is crucial \cite{perez2023simulating, schnappinger2024disentangling}. In the lowest order, collective vibrational excitations involve two excitations localized either within the same or across different vibrational modes. When one- and two-quantum cavity excitations are externally manipulated via optical driving, the resulting cavity-dressed vibrational interactions and anharmonicities can be probed. These interactions directly affect the properties of the vibrational excited states. Previous spectroscopic studies deployed time-resolved, two-pulse pump-probe techniques to track the cavity-dressed vibrational dynamics \cite{stemo2024ultrafast, saurabh2016two}. When higher-order vibrational excitations are present, they may not unambiguously resolve the dynamics. Three-pulse stimulated spectroscopies can selectively track transitions between the ground and first excited states, as well as between the first and second excited manifolds. They are particularly effective when multiple excitation manifolds and dissipation are present. Throughout this article, the bare and cavity-dressed vibrational modes are referred to as vibrons and vibron-polaritons, respectively. The auxiliary vibrational modes, which effectively act as a reservoir, are referred to as phonons.\\
In this work, we extend previous investigations in three principal ways. First, we present a theoretical framework for cavity-dressed interacting vibrons that includes both one- and two-vibron states, thereby accounting for their collective anharmonic nature, specifically overtone and combination vibrational nonlinearities, in a consistent manner. Second, we expand the microscopic theory to include dissipation by accounting for both low- and high-frequency reservoir modes. Third, we deploy two three-pulse stimulated spectroscopic techniques, photon echo (PE) and double-quantum coherence (DQC) spectroscopy, to investigate the dissipative dynamics of one- and two-vibron-polaritons. 
The spectroscopic techniques presented here form a multimodal measurement framework in which two complementary methods track the same spectral modes to unravel their participation in two distinct types of dynamics. We focus on a parameter regime where the cavity–vibron coupling strengths are comparable to those between the vibrons. This regime allows for the modulation of vibrational excited-state delocalization without fundamentally altering the energy gaps. It is essential for controlling of vibrational energy redistribution while maintaining the spectroscopic identity of the vibrons.\\
The article is organized as follows: section~\ref{sec:dissonetwo} is divided into four subsections. Section~\ref{subsec:vibpolham} develops the model for interacting, dissipative vibron-polaritons and their coupling to phonons. Section~\ref{subsec:vibpoltensor} discusses the vibron-polariton tensor-product basis, the resulting eigenbasis, and the simulation parameters. The polariton Green's functions, which describe the nonlinear response, are detailed in section~\ref{subsec:vibpolgf}. Finally, the external laser driving is discussed in section~\ref{subsec:vibpollaser}. Appendices~\ref{app:ham} and \ref{app:gme} complement these discussions.
Section~\ref{sec:mdcs} is divided into two subsections: section~\ref{subsec:dqc} and section~\ref{subsec:pe} present the mathematical expressions for double-quantum coherence and photon-echo spectroscopy, respectively, and discuss their numerical simulations. The corresponding derivations for these signals are provided in appendices~\ref{app:response0} and \ref{app:response}.
Section~\ref{sec:conclusion} summarizes the results, highlights the advantages of the investigated parameter regime and possible limitations. It also discusses potential future directions.
\begin{figure*}[ht!]
\begin{center}
\includegraphics[width=.96\textwidth]{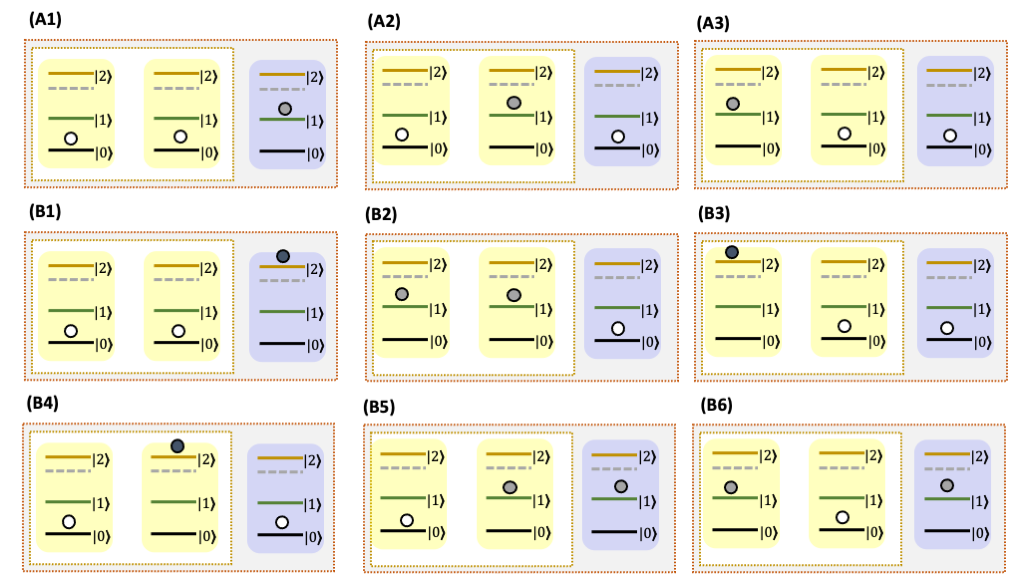}
\end{center}
\caption{Illustration of cavity-vibron configurations. The first row (A1–A3) represents single-excitation exchange processes, whereas the second and third rows (B1–B6) represent two-excitation processes. The gray background represents configuration mixing facilitated by cavity-vibron coupling, while the white background indicates mixing via vibron hopping.
The ground states correspond to the absence of excitations. The one-polariton states comprise weighted contributions from three configurations: the pure one-photon cavity ( $\ket{00}\ket{1}_c$) and the mixed cavity-vibron configurations ($\ket{10}\ket{0}_c, \ket{01}\ket{0}_c$)the first row. The two-polariton states comprise weighted contributions from six configurations: the pure two-vibron states ($\ket{20}\ket{0}_c, \ket{02}\ket{0}_c, \ket{11}\ket{0}_c$), the pure two-photon cavity state ($\ket{00}\ket{2}_c$), and the mixed cavity-vibron configurations ($\ket{10}\ket{1}_c, \ket{01}\ket{1}_c$).}
\label{fig:vibpoleet}
\end{figure*}
\section{Dissipative one- and two-polariton kinetics}
\label{sec:dissonetwo}
\subsection{Vibron-polariton Hamiltonian}\label{subsec:vibpolham}
A model for interacting vibrons can be derived from the ab-initio Hamiltonian (described in Appendix~\ref{app:ham}), which leads to the cavity-vibron-phonon Hamiltonian (with $\hbar=1$)
\begin{align}
    H_p  &=
\sum_{m} E_{m}^{} B_{m}^\dag B_{m}^{} +\sum_{mn} J_{mn}^{} B_{m}^\dag B_{n}^{} \nn\\&
+\sum_{m} U_{mm}^{(1)} B_{m}^\dag B_{m}^\dag B_{m}^{} B_{m}^{}+ \sum_{mn} U_{mn}^{(2)} B_{m}^\dag B_{n}^\dag B_{m}^{} B_{n}^{} \nn\\
&+\sum_m \sum_{k_c}^{} \omega_{k_c}^{(c)} a_{k_c}^\dag a_{k_c}^{} + \sum_{m,k_c} g_{m k_c}^{(c)}(a_{k_c}^\dag B_m^{} + B^\dag_m  a_{k_c}^{} ) \nonumber\\
&+\sum_m \sum_{k}^{} \omega_k^{(b)} b_k^\dag b_k^{} +\sum_{m,k} g_{mk}^{(b)} B_{m}^\dag B_{m}^{} (b_{k}^{}+b_k^\dag)
\label{eqn:ham}
\end{align}
In the following, we discuss the Hamiltonian in detail.\\
\emph{Interacting vibrons---}
The first two terms of the Hamiltonian represent interacting vibron modes with excitation energies $E_{m}$ and inter-mode couplings $J_{mn}$. These coupling magnitudes can be estimated from the transition dipole matrix elements, inter-mode distances, and relative orientations. 
Cooperative vibrational excitations can be classified as either combination or overtone modes, referring to inter-mode and intra-mode two-vibron excitations, respectively. The third and fourth terms of the Hamiltonian represent the nonlinearities of the overtone and combination modes, denoted by $U_{mm}^{(1)}$ and $U_{mn}^{(2)}$, respectively. These parameters are related to the anharmonicities of the vibron modes and affect the two-vibron Hamiltonian matrix elements: $H_{mnkl}^{(2)} = H_{mn}^{(1)} \delta_{kl} + \delta_{mn} H_{kl}^{(1)}$, where the one-vibron Hamiltonian matrix elements are $H_{mn}^{(1)}=\sum_{mn} E_{m}^{} \delta_{mn} + J_{mn}^{} B_{m}^\dag B_{n}^{}$.
The magnitudes of vibron nonlinearities depend on the nature of the permanent dipole matrix elements. 
Together with the inter-mode coupling $J_{mn}$, the vibron nonlinearities govern the extent of eigenstate delocalization in vibrons. The vibron creation (annihilation) operators, $B_{m}^{\dag}$ ($B_{m}$), follow the commutation relations $[B_{m}^{\dag},B_{n}]=\delta_{mn}$ and
$[B_{m},B_{n}]=[B_{m}^{\dag},B_{n}^{\dag}]=0$.\\
\emph{Cavity---}The fifth and sixth terms in the Hamiltonian describe the cavity and the cavity-vibron coupling, respectively. The latter is governed by the parameter $g_{m k_c}^{(c)}$. In this work, we consider a single cavity mode ($k_c=1$) with frequency $\omega_{k_c}^{(c)}$ that hosts both one-quantum and two-quantum cavity excitations. We operate in the idealized, narrowband limit characterized by low cavity decay rates, assuming the cavity mode is pumped by a low-intensity laser source. \\
In a Distributed Bragg Reflector (DBR) Fabry–Perot cavity, the cavity mode energies depend on the angle of incidence, $\theta$, of the driving laser source and the refractive index of the medium, $n_{\text{ind}}$. These quantities are related via the following expression: $\omega_c = \omega_0 /(\sqrt{1-\sin^2{\theta}/n_{\text{ind}}^2})$.
For DBR cavities, semi-empirical coupling strengths are determined by the relation: $g_{m k_c}^{(c)} = \sqrt{N} d_c \cdot E_c \sqrt{\omega_c/2 \varepsilon_0 V}$, where $N$ where $N$ is the number of molecules interacting with the mode $c$, $d_c (E_c)$ represent the transition dipole moment (the cavity mode electric field vector), respectively. 
The mode volume $V_c=\lambda/n_{\text{ind}}$ is characterized by the resonant wavelength $\lambda$ and the effective intra-cavity refractive index $/n_{\text{ind}}$.\\
\emph{Phonons---}
The seventh and eighth terms of the Hamiltonian represent the phonon and vibron–phonon interaction terms, respectively. The phonon creation (annihilation) operators, $b_{k}^{\dag}$ ($b_{k}$), with frequencies $\omega_{k}^{(b)}$, follow the free-boson commutation relations: $\comm{b_j}{b_{j'}} =\comm{b_j^\dagger}{b_{j'}^\dagger}=0$ and $\comm{b_j^{}}{b_{j'}^\dagger} = \delta_{jj'}^{}$. 
Each vibron mode is assumed to be independently coupled to a set of non-interacting phonon modes, characterized by the coupling strengths $g_{mk}^{(b)}=g_{m}^{(b)}g_{k}^{(b)}$. We can estimate the phonon spectral function from the distribution of these coupling strengths, expressed as:
\begin{align}
    C_{}(\omega) = \pi\sum_j |g_{k}^{(b)}|^2\Big(\delta(\omega-\omega_j) +\delta(\omega+\omega_j)\Big)
\end{align}
In the expression above, $\omega_{j}$ and $g_{k}^{(b)} = \sqrt{S_{j}} \omega_{j}$ represent the site-independent phonon mode frequencies and couplings, respectively (where $S_{j}$ is the Huang-Rhys factor).
We describe the phonon spectral function using one overdamped Brownian oscillator and one structured Brownian oscillator. In the limit of a continuous frequency distribution, the spectral function takes the form:
\begin{align}\label{eqn:phononspecf}
J(\omega) &= \frac{2\lambda_0 \,\gamma_0 \omega}{(\omega^2+\gamma_0^2)}+ \frac{2\lambda_{} \,\upsilon_{}^2  \gamma_{} \omega}{((\upsilon_{}^2-\omega^2)^2+\omega_{}^2\gamma_{}^2) }
\end{align}
The overdamped Brownian oscillator model provides an accurate description of the low-energy phonon modes characteristic of molecular solvents. In comparison, the structured Brownian oscillator describes high-frequency vibrations \cite{mukamel1995principles, cho2009two, debnath2020entangled, debnath2022entangled}. As demonstrated later, these phonon modes act as finite-bandwidth reservoirs that accommodate all relevant dissipative phenomena needed for this work, namely, intra-manifold dephasing, transport, and inter-manifold dephasing.
\begin{figure}[ht!]
\begin{center}
\includegraphics[width=.48\textwidth]{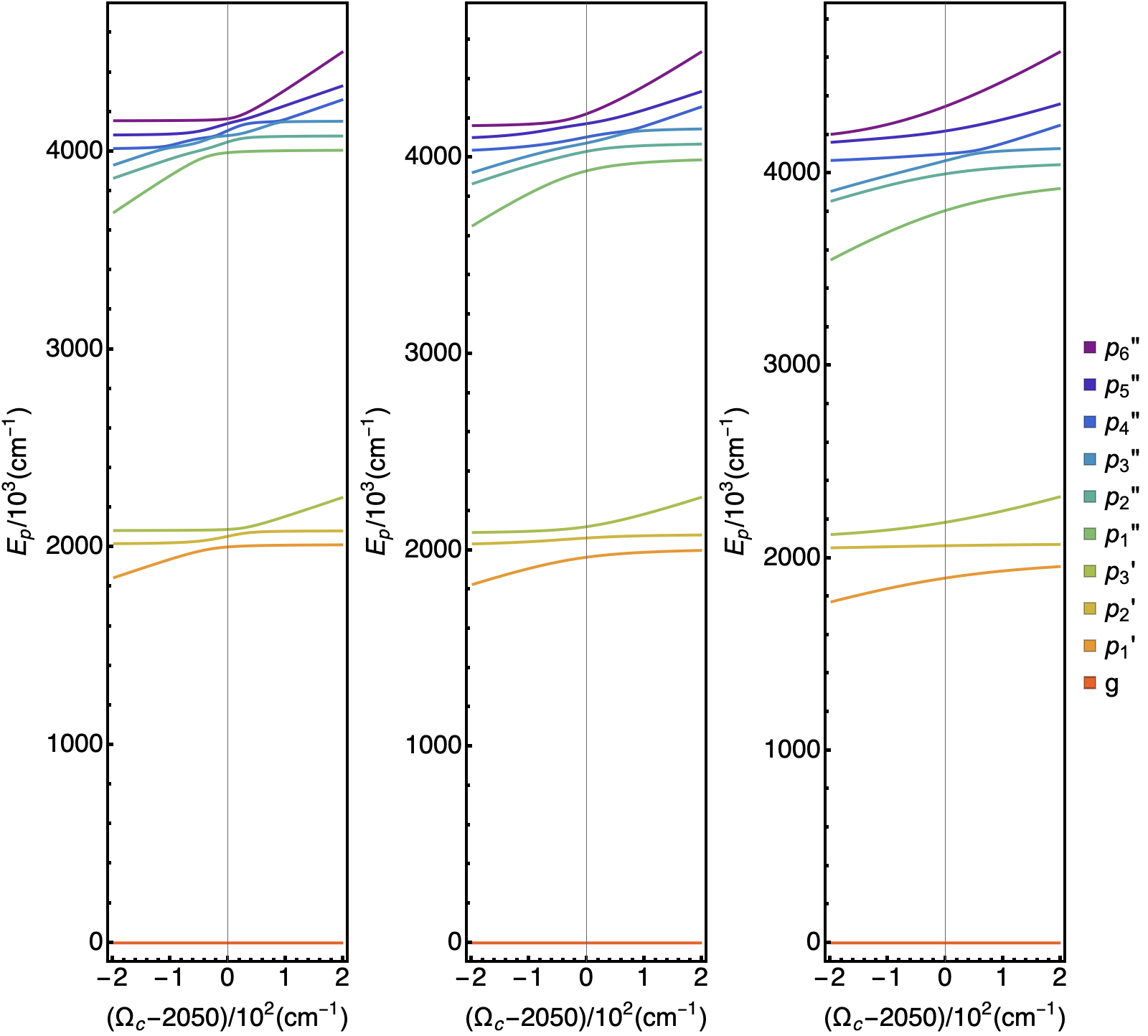}
\end{center}
\caption{Parametric dependence of one- and two-polariton energies on the cavity mode detuning (defined via the scanning parameter $\Omega_c)$ for three distinct coupling strengths. The coupling strength increases from left to right, resulting in a marked increase in state repulsion. For any combination of cavity detuning and coupling strength, a specific energetic ordering of states is obtained. By selecting appropriate laser parameters, the time-dependent fluctuations of the polariton energy gaps can be monitored. Notably, the relaxation parameters are also determined by these gaps.}
\label{fig:vibpolenergy}
\end{figure}
\subsection{Vibron-polariton states}\label{subsec:vibpoltensor}
\emph{Vibron-polariton states tensor product basis---}
The vibron-polariton states can be constructed using a tensor-product basis. These basis states consist of the direct product between local vibrational modes and the cavity mode, $\ket{n_{v_2} n_{v_1}n_{c}}=\ket{n_{v_2} n_{v_1}}\ket{n_{c}}_c$, where the $n_v (n_c)$ denote the occupation numbers of the vibrational and cavity modes, respectively. They represent various configurations.
Since the physical processes relevant for the DQC and PE techniques are restricted to one- and two-polariton manifolds, the configuration space can be effectively truncated. \\
The one-polariton basis set is spanned by three configurations: $\ket{10}\ket{0}_c$, $\ket{01}\ket{0}_c$, and $\ket{00}\ket{1}_c$. The first two configurations represent pure vibrational excitations, whereas the last represents a pure cavity excitation. 
The cavity-vibron interaction Hamiltonian mediates transitions between the configurations $\ket{01}\ket{0}_c$ and $\ket{10}\ket{0}_c$ via the intermediate $\ket{00}\ket{1}_c$. The two-polariton basis set is spanned by six configurations. They are pure two-vibron excitations: $\ket{20}\ket{0}_c, \ket{02}\ket{0}_c, \ket{11}\ket{0}_c$, pure two-photon cavity excitation: $\ket{00}\ket{2}_c$, and and mixed one-photon–one-vibron excitations: $\ket{10}\ket{1}_c, \ket{01}\ket{1}_c$. 
The cavity-vibron interaction Hamiltonian mediates transitions between the pure two-vibron and pure two-photon configurations exclusively through the mixed cavity-vibron states. For example, the transitions between $\ket{20}\ket{0}_c$ and $\ket{02}\ket{0}_c$ occur via the intermediate $\ket{10}\ket{1}_c$ or $\ket{01}\ket{1}_c$. The complete suite of configurations is displayed in Fig.~\ref{fig:vibpoleet}.\\
\begin{figure*}[ht!]
\begin{center}
\includegraphics[width=.96\textwidth]{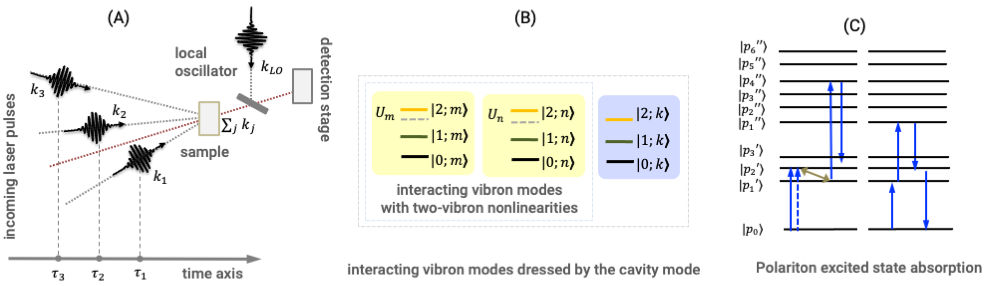}
\end{center}
\caption{(A) Schematic diagram of the pulse configuration for multidimensional coherent spectroscopy, including the heterodyning stage (adapted from \cite{hochstrasser2007two}). For clarity, relative dimensions are exaggerated. The sample consists of a cavity-encapsulated system of vibrons, polaritons, and phonons. (B) The cavity mode interacts with two interacting vibron modes. Each has distinct two-vibron nonlinearities, illustrated in the local mode description. C) Two plausible dynamical pathways (in Albrecht notation) displaying polariton excited-state absorption, depicted in the polariton eigenbasis. The pathways on the left are pertinent to photon echo spectroscopy, while those on the right are pertinent to double-quantum coherence spectroscopy.}
\label{fig:vibpoltech}
\end{figure*}
\emph{Vibron-polariton eigenbasis---}
The spectroscopic signals, however, are best interpreted in the driving-field-free eigenbasis. We obtain the polariton eigenstates, $\ket{p_0}, \ket{p'}, \ket{p''}$ via exact diagonalization of the first six terms of the Hamiltonian in Eq.~\ref{eqn:ham}. The procedure yields $n_0=1$ ground state ($\ket{p_0}$), $n_{p'}=3$ one-polariton states ($\ket{p'}$), and $n_{p''}=6$ two-polariton states ($\ket{p''}$) distributed over three number-conserving manifolds. These states can be expressed as:
\begin{align}
\ket{p_0} &= \ket{g}\\
\ket{p'} &= \sum_{m k_c} \big(\phi_{mp'}^{(1)} B_s^\dag +  \phi_{k_c p'}^{(1)} a_{k_c}^\dag\big)\ket{g}\\
\ket{p''} &= \sum_{mn k_{c,i} k_{c,j}} \big(\phi_{mnp''}^{(2)} B_m^\dag B_n^\dag + \phi_{m k_{c,i} p''}^{(2)}  B_m^\dag a_{k_{c,i}}^\dag \nn\\&
+ \phi_{ k_{c,i} k_{c,j} p''}^{(2)} a_{k_{c,i}}^\dag a_{k_{c,j}}^\dag \big)\ket{g}
\label{eqn:site2eig}
\end{align}
One may use the transformation matrix elements $\phi_{ab}^{(n)}$ to estimate the degree of eigenstate delocalization over the tensor-product states. The exact diagonalization procedure incorporates the effects of hopping and cavity–vibron coupling non-perturbatively.\\
The progression of polariton energy levels as a function of cavity frequency is presented in Fig.~\ref{fig:vibpolenergy} for three distinct coupling strengths. 
In these plots, we observe a state repulsion within each manifold, a characteristic feature which is a joint function of cavity coupling strengths, hopping, and cavity detuning. Selecting a specific cavity frequency fixes the relative ordering of the energy levels for a given coupling strength. Because nonlinear spectroscopic techniques probe time-dependent fluctuating energy gaps, this figure may serve as a guide for selecting the laser parameters to be deployed.\\
\emph{Simulation parameters---} 
For the numerical simulations, we use parameters that closely resemble those of the carbonyl stretching vibrations in metal carbonyl complexes, such as $\text{Rh(CO)}_2 \text{C}_5 \text{H}_7 \text{O}_2$ in $\text{RDC}$. The vibron mode energies are $E_1 = 2019 ~\text{cm}^{-1}$ and $E_2 = 2080~\text{cm}^{-1}$, while the hopping energies are $J_{12}=J_{21}= -16~\text{cm}^{-1}$. The nonlinearity parameters are specified as $U_{11}^{(2)}= -11~\text{cm}^{-1}$, $U_{22}^{(2)}= -14~\text{cm}^{-1}$, and $U_{12}^{(2)}= -14~\text{cm}^{-1}$. 
Depending on the magnitude and sign of the nonlinearity parameters, the energetics and dynamics of cooperative two-vibron states differ significantly from those of the constituent two non-interacting one-vibron excitations. A negative sign, for example, represents the stabilization of the two-vibron states. These vibron nonlinearity parameters contribute to the nonlinear response even in the absence of cavity interactions.  
The cavity mode energy is, $E_c = 2050~\text{cm}^{-1}$. It is energetically positioned between the vibron modes, specifically, it is slightly detuned from both modes. This is a deliberate choice, considering the narrowband nature of the cavity mode and the intended dispersive coupling regime \cite{gonzalez2024light, baraillon2020linear, yu2019two}. The mode couples uniformly to both vibron modes with coupling strengths $g_{1,k_c}^{(c)}=g_{2,k_c}^{(c)}= 25~\text{cm}^{-1}$. 
The cavity coupling strengths in this work are comparable to the hopping energy scales (i.e., $J_{mn} \sim g_{k_c}^{(c)}$) and the mode is assumed to exhibit no spatial dependence at the molecular scale. 
For the phonon modes, we use the following parameters: $\lambda_0 = 15 \text{cm}^{-1}$, $\gamma_0 = 20 \text{cm}^{-1}$ for the overdamped oscillator, and $\lambda = 15 \text{cm}^{-1}$, $\upsilon = 721 \text{cm}^{-1}$ for the structured oscillator. The variables $\gamma_{0}^{-1}, \upsilon_{}^{-1}$ define the timescales of the relaxation dynamics and can be estimated from linear absorption and fluorescence signals \cite{mukamel1995principles}.
To illustrate the discriminatory role of relaxation, we maintained the vibron–phonon coupling strengths for the two modes at a relative ratio of $g_{2,k}^{(b)}/g_{1,k}^{(b)} =1.25$.\\
Using this set of parameters, we obtain three one-polariton states with energies (in the units of $\text{cm}^{-1}$): $E_{p_1'}=1996.59, E_{p_2'}=2058.21, E_{p_3'}=2094.2$, and six two-polariton states, with energies: $E_{p_1''}=4011.45, E_{p_2''}=4030.4, E_{p_3''}=4088.37, E_{p_4''}=4111.39, E_{p_5''}=4149.21, E_{p_6''}=4165.19 $, (expressed in the unit of $\text{cm}^{-1}$). Using these values, the polariton energy gaps, $\omega_{p_j' p_k''}$ can be evaluated. 
We list the frequencies here for convenience: $\omega_{p_1' p_k''} \equiv 2014.85, 2033.80, 2091.78, 2114.79, 2152.62, 2168.59$
$\omega_{p_2' p_k''} \equiv 1953.24, 1972.19, 2030.17, 2053.18, 2091.00, 2106.98$ 
$\omega_{p_3' p_k''} \equiv 1917.25, 1936.20, 1994.18, 2017.19, 2055.01, 2070.99$. Here, the index $k$ denotes the two-polariton states ordered by increasing energy.
\subsection{Vibron-polariton Green's functions}\label{subsec:vibpolgf}
Throughout this work, we express the nonlinear response functions using phonon-averaged polariton Green's functions. The poles of the polariton Green's functions are determined by the polariton energy gaps and their associated broadening. The latter can be estimated using the microscopic models of polariton dissipation. In particular, the phonon modes drive three distinct mechanisms: intra-manifold transport, as well as inter- and intra-manifold dephasing. In Appendix~\ref{app:gme}, we describe the techniques to obtain the relaxation parameters corresponding to each mechanism starting from the Markovian relaxation kernel. \\
The matrix elements of the Green's function describing intra-manifold transport and dephasing can be expressed as:
\begin{align}
    &G_{p_1 p_2,p_3 p_4}^{(N)}(t) = \delta_{p_1 p_2}\delta_{p_3 p_4} \theta(t) [\exp{(- K^{} t)}]_{p_1 p_1,p_3 p_3} \nn\\
&+ (1-\delta_{p_1 p_2}) \delta_{p_1 p_3}\delta_{p_2 p_4} \theta(t) \exp{(-i \omega_{p_1 p_3}^{} t -\gamma_{p_1 p_3}^{(N)} t)}
\label{eqn:gf}
\end{align}
Here, the second term describes the polariton dephasing of $p_1p_3$ states, whereas the first term describes polariton population transport between $p_3p_3$ and $p_1p_1$ states. The latter is useful for developing an intuitive picture of how the population may evolve following state-specific excitations. We present the simulation of population evolutions, starting from each of the one- and two-polariton states, in Fig.~\ref{fig:vibpolpopone} and Fig.~\ref{fig:vibpolpoptwo}, respectively. 
The matrix elements of the Green's function describing intra-manifold coherence are given by
\begin{align}\label{eqn:gfdephasinginter}
    G_{p_1'p_2,p_1'p_2}(t) &= \exp{[-i\omega_{p_1'p_2} t-\gamma_{p_1' p_2^{}}^{} t]}
\end{align}
where primed and non-primed indices in $p_1'p_3$ denote states belonging to different manifolds.\\
The two key quantities that govern the dissipation parameters are the polariton overlap function and the relaxation function (see Appendix~\ref{app:gme}). In the above, we neglected phonon-induced spectral shifts and the effects of driving-induced dissipation \cite{debnath2012chirped, debnath2013high, debnath2013dynamics}.
\begin{figure*}[ht!]
\begin{center}
\includegraphics[width=.96\textwidth]{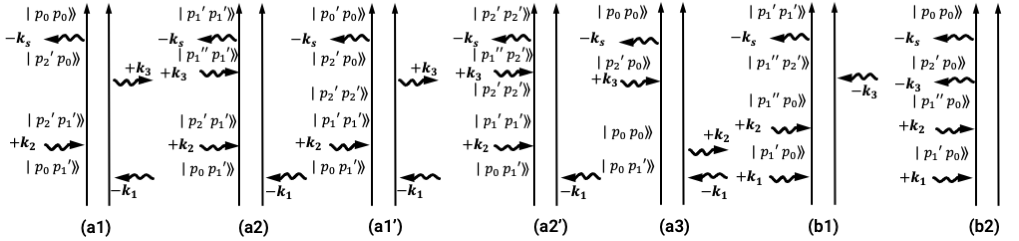}
\end{center}
\caption{Displayed are the Liouville space Feynman diagrams that describe the polariton pathways contributing to the PE and DQC signal. The time evolution of the polariton density operator, denoted $\ket{ab}\rangle \rightarrow \ket{a}\bra{b}$, is depicted by vertical pairs of arrows. Wiggly lines pointing toward or away from the density operator indicate the resonant absorption or emission of a single excitation, respectively. Diagrams labeled (a1)–(a3) denote photon-echo pathways, while (b1)–(b2) denote double-quantum coherence pathways. }
\label{fig:vibpolfeyn}
\end{figure*}
\subsection{Vibron-polariton-laser interactions}\label{subsec:vibpollaser}
The nonlinear response of the dissipative polariton modes is generated by the interaction with three incoming laser pulses. Within the rotating wave approximation (RWA), the laser–polariton interaction Hamiltonian is given by:
\begin{align}
&H_{\text{int}} = \sum_{m}  E(t) \left( d_{m}^{}a_{k}^\dag + \sqrt{2} d_{m}^{} a_{k}^\dag a_{k} a_{k} + \text{h.c} \right)
\end{align}
The incoming laser field, consisting of a sequence of non-overlapping laser pulses: 
\begin{align}
    E(t) &=\sum_{j}\mathcal{E}_{j}^{}(t-\tau_{j})\exp[i(\bm{k}_{j} \bm{r}-\omega_{j}(t-\tau_{j}))]+ \text{c.c.}
\end{align}
and $d_{m}^{} $ is the transition dipole operator, $\omega_j$ is the frequency, $\mathbf{k}_j$ is the wavevector, and $\tau_j$ is the center-of-arrival time for each pulse. The temporal centering is crucial for establishing the specific pulse sequence ordering required for nonlinear spectroscopies with finite pulse shapes. Detailed derivations are provided in Appendix~\ref{app:response0}.\\
The external driving operates in a cavity-pumping configuration, creating excitations via the cavity mode operator. As discussed in the section~\ref{subsec:vibpoltensor}, and illustrated in Fig.~\ref{fig:vibpoleet}, the laser driving promotes transitions between configurations that differ by one cavity quantum. Individual transitions can be manipulated by adjusting the spectral bandwidths of the pulses.
The interaction Hamiltonian is transformed into the eigenbasis using the transformation matrices defined in Eq.~(\ref{eqn:site2eig}), yielding renormalized transition dipole matrix elements.
This assumption of weak multi-pulse driving offers several advantages: the low mean energy pumped into the polariton modes avoids unwanted laser-induced nonlinearities, driving-induced dissipation is minimized, and the signal scales favorably with the intensity of the incoming fields \cite{zhou2024nature, debnath2012chirped, debnath2013dynamics}.
\section{Coherent multidimensional spectroscopy}
\label{sec:mdcs}
In this section, we present the theory and numerical simulations for the double-quantum coherence (DQC) and photon echo (PE) signals. Detailed derivations are provided in Appendices~\ref{app:response0}–\ref{app:response}.
The double-quantum coherence (DQC) signal, selected in the phase-matching direction $\mathbf{k}_{\text{III}} = \mathbf{k}_1 + \mathbf{k}_2 - \mathbf{k}_3$, is sensitive to polariton excited-state absorption ans stimulated emission processes. This signal probes resonances that originate from polariton mode correlations, which are, in turn, related to the cavity-modulated vibron nonlinearity terms. 
The photon-echo (PE) signal, selected in the phase-matching direction $\mathbf{k}_{\text{I}} = -\mathbf{k}_1 + \mathbf{k}_2 + \mathbf{k}_3$, is sensitive to polariton ground-state bleaching, excited-state absorption, and stimulated emission processes. The signal predominantly probes resonances arising from cavity-modulated vibron mode coupling and transport kinetics. \\
We note that choosing a specific MDCS technique fixes the wavevectors ($\mathbf{k}_j$) and the pulse-centering times ($\tau_j$, or more specifically, the time delays $T_j$). Thus, in the simulation, $\tau_{0,j}$ and $\omega_j$ are the only parametric variables to be varied while computing the spectra. 
In the simulation, the laser field profiles are taken as Gaussian $\mathcal{E}_j(\omega)= \sqrt{\pi/\Gamma_{0,j}}\exp{[-(\omega-\omega_{0,j})^2/ 4 \Gamma_{0,j}]}$, where $\Gamma_{0,j}$ is related to the temporal width of the pulse $\tau_{0,j}$ via $\Gamma_{0,j} = 2\mathrm{ln}(2)/\tau_{0,j}$.
\subsection{Double-quantum coherence spectroscopy}
\label{subsec:dqc}
\subsubsection{Theory}\label{subsubsec:dqctheory}
During DQC signal generation, the first two laser–polariton interactions induce a time-dependent polarization, dominated by two-polariton coherence. It evolves over a parametric time interval before being projected onto two discriminating one-polariton coherence components. The differential nature of these two pathways, largely determined by the last interfering polarization components, dictates the magnitude of the signal. The corresponding pathways, denoted $\text{b1}-\text{b2}$, are depicted in Fig.~\ref{fig:vibpolfeyn}.
Following the derivations outlined in Appendices~\ref{app:response0} and \ref{app:response}, the DQC signal is expressed as:
\begin{widetext}
\begin{align}
& S_{\mathrm{DQC}}(\Omega_2,\Omega_3)=  i^3 c_0\times\sum_{p_{2}'' p_{2}' p_{1}' p_{0}} \Big\{
\frac{
d_{p_{1}'' p_{2}'}\mathcal{E}_s^{*}(\omega_{p_{1}'' p_{2}'}-\omega_s) 
d_{p_{2}' p_{0}^{}}\mathcal{E}_3^{*}(\omega_{p_{2}' p_{0}^{}}-\omega_3)  d_{p_{1}''p_{1}'}\mathcal{E}_2(\omega_{p_{1}''p_{1}'}-\omega_2) 
d_{p_{1}' p_{0}}\mathcal{E}_1(\omega_{p_{1}' p_{0}}-\omega_1)} 
{(\Omega_3-z_{p_{2}'' p_{2}'}) (\Omega_2-z_{p_{2}'' p_{0}})} 
\nonumber\\&
\times \exp(-i z_{p_{1}'p_{0}}t_{1}) - \frac{
d_{p_{1}'' p_{1}'}\mathcal{E}_s^{*}(\omega_{p_{1}'' p_{1}'}-\omega_s)
d_{p_{2}' p_{0}}\mathcal{E}_3^{*}(\omega_{p_{2}' p_{0}}-\omega_3) d_{p_{1}''p_{1}'}\mathcal{E}_2(\omega_{p_{1}''p_{1}'}-\omega_2) 
d_{p_{1}' p_{0}}\mathcal{E}_1(\omega_{p_{1}' p_{0}}-\omega_1)} 
{(\Omega_3-z_{p_{2}' p_{0}}) (\Omega_2-z_{p_{1}'' p_{0}})} \exp(-i z_{p_{1}'p_{0}}t_{1})
\Big\}  
\label{eqn:dqc}
\end{align}
\end{widetext}
The two-dimensional DQC spectra can be displayed using two scanning frequencies, $\Omega_2$ and $\Omega_3$, for fixed parametric values of $t_1$. The associated peaks along $\Omega_2$ correspond to the two-polariton resonances $z_{p_2' p_0}$, while those along $\Omega_3$ correspond to the one-polariton resonances $z_{p_1' p_0}$ or $z_{p_2' p_1'}$.
\subsubsection{Simulations}\label{subsubsec:dqcsim}
\begin{figure*}[ht!]
\begin{center}
\includegraphics[width=.96\textwidth]{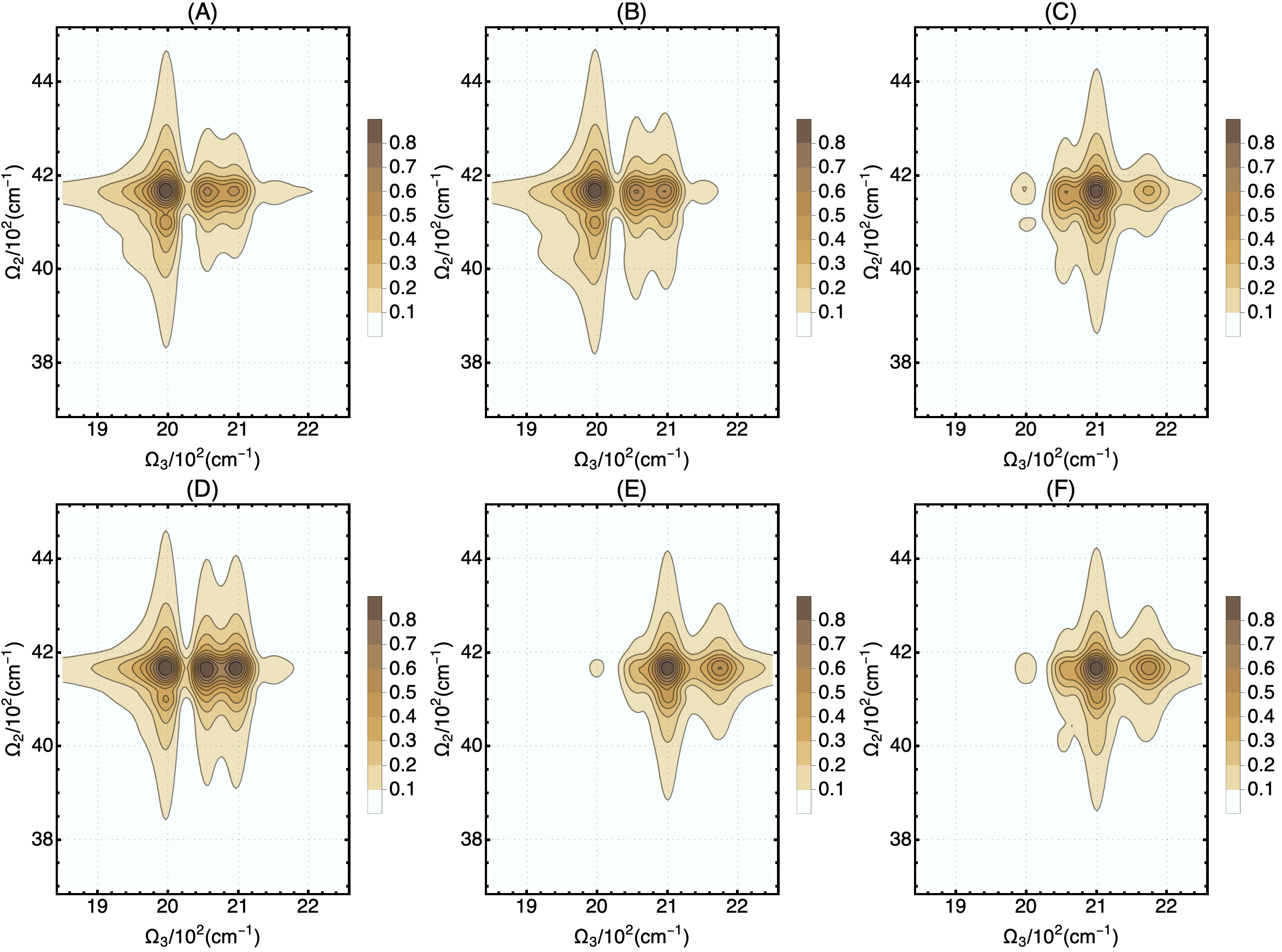}
\end{center}
\caption{Simulated two-dimensional double-quantum coherence (DQC) spectra presented for six variations of the driving laser parameters. These variations explore the temporal width of the excitation pulses, the excitation frequencies, and the final projection frequencies. Off-diagonal spectral features represent correlations between the one-polariton resonances. A careful selection of pulse parameters can amplify or suppress particular features which shows the accumulation of spectral weight in specific resonances. The expression in Eq.~\ref{eqn:dqc} and the pathways depicted in Fig.~\ref{fig:vibpolfeyn} rationalize the observed spectral features. For a detailed discussion, see section~\ref{subsubsec:dqcsim}.} 
\label{fig:vibpoldqc}
\end{figure*}

The numerical simulations are presented in Fig.~\ref{fig:vibpoldqc}, denoted (A)-(F). The first delay time is set to $t_1 = 0$, which implies that the initial two laser–polariton interactions are coincident.
For the reference case presented in (A), the temporal widths of the incoming and heterodyning laser pulses are $\tau_1 = \tau_2 = \tau_3 = 100$ fs and $\tau_4 = 100$ fs, respectively. The central frequencies of these pulses are selected to coincide with specific polariton transitions by setting $\omega_1 = \omega_2 = \omega_{p_2'p_0}^{}$ and $\omega_3 = \omega_4 = \omega_{p_5''p_1'}^{}$.
This configuration ensures that the final two interactions preferentially excite a specific energy sector, the higher energy region in this instance. Throughout the following sections, values of $\tilde{\Omega}_j$ are quoted in units of $10^2$ cm$^{-1}$.
We observed three principal resonances along $\tilde{\Omega}_3$, denoted by $(\tilde{\Omega}_{3,a}, \tilde{\Omega}_{3,b}, \tilde{\Omega}_{3,c}) \approx (20.0, 20.7, 21.0)$, and two along $\tilde{\Omega}_2$, denoted by $(\tilde{\Omega}_{2,a}, \tilde{\Omega}_{2,b}) \approx (41.0, 41.5)$. Peaks along the $\tilde{\Omega}_2$ axis exhibit larger spectral broadening than those along $\tilde{\Omega}_3$. The off-diagonal peaks represent polariton-polariton correlations originating from cavity-dressed two-vibron nonlinearity terms. The subsequent results incorporate parametric variations of this reference simulation.\\
In (B), we investigate the role of narrowband laser excitation by setting $\tau_0 = 40$ fs. The signal reveals a minor spectral narrowing of the peaks at $(\tilde{\Omega}_{2,b},\tilde{\Omega}_{3,b})$ and $(\tilde{\Omega}_{2,b},\tilde{\Omega}_{3,c})$. However, the broader correlation features remain relatively unaffected. This observation indicates two features of the dynamics: the dephasing dominates the inter-manifold polariton dynamics and the two-polariton coherence, despite the higher density of states remain nonselective. \\
To assess these observations further, (C) presents results using different projection frequencies by setting $\omega_3 = \omega_4 = \omega_{p_2''p_1'}^{}$, while maintaining the other parameters from (A). These results differ markedly from the previous two cases. The peak at $(\tilde{\Omega}_{2,b},\tilde{\Omega}_{3,b})$ is narrower and exhibits reduced spectral weight. In contrast, $(\tilde{\Omega}_{2,b},\tilde{\Omega}_{3,c})$ is prominent and broader, and a new peak $(\tilde{\Omega}_{2,b},\tilde{\Omega}_{3,d})$ appears at $\tilde{\Omega}_{3,d} \approx 20.8$. The change in the final-state projections and the higher density of states in the two polariton manifold has resulted in the observations along $\tilde{\Omega}_3$. \\
In (D), we investigate the effects of different state initializations by setting the central frequencies of the excitation pulses to $\omega_1 = \omega_2 = \omega_{p_3'p_0}^{}$. Compared with (A), the spectral weight is shifted to peaks $(\tilde{\Omega}_{2,b},\tilde{\Omega}_{3,b})$ and $(\tilde{\Omega}_{2,b},\tilde{\Omega}_{3,c})$. They also exhibit peak-broadening. The peak at $(\tilde{\Omega}_{2,a},\tilde{\Omega}_{3,a})$ becomes narrower. We note that the targeted state is energetically the highest in the one-polariton manifold. The spectral changes indicate that the excitation conditions altered the excited-state absorption pathways.\\
In (E), we alter both the initial excitation and final projection frequencies to assess the relative roles of the initial and final laser-polariton interactions. We observed two distinct features: the projection stage dominates the response, and the peak at $(\tilde{\Omega}_{2,b},\tilde{\Omega}_{3,b})$ vanishes. Consequently, the spectral weights are transferred to the remaining peaks.\\
Finally, in (F), we explore the effect of a narrowband pulse on the simulation shown in (E). We observed broadening of the off-diagonal peak at $(\tilde{\Omega}_{2,b},\tilde{\Omega}_{3,d})$. This suggests that the initial state excitation and projection are not affected by the temporal width of the pulses. 
\begin{figure*}[ht!]
\begin{center}
\includegraphics[width=.96\textwidth]{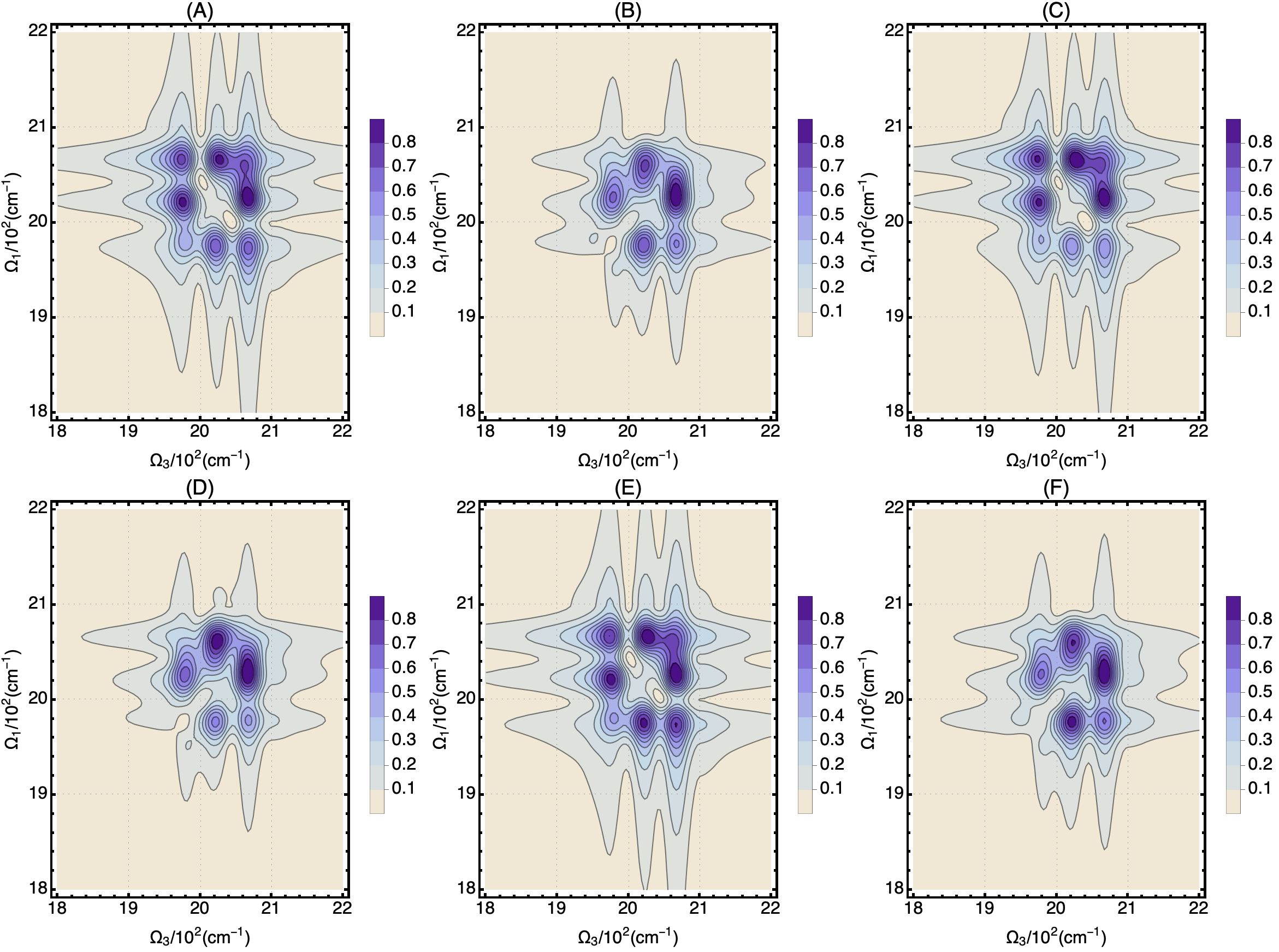}
\end{center}
\caption{Simulated two-dimensional photon-echo spectra are presented for six variations of the driving laser parameters. These variations explore broadband excitation of one-polariton states, dependence on transport and effect of narrowband excitations. Off-diagonal spectral features provide information regarding the polariton mode couplings induced by transport, typically interpreted in terms of one-polariton resonances. The expression in Eq.~\ref{eqn:pe} and the pathways depicted in Fig.~\ref{fig:vibpolfeyn} can be used to rationalize the spectra. For detailed discussions see section~\ref{subsubsec:pesim}.}
\label{fig:vibpolpe}
\end{figure*}
\subsection{Photon-echo spectroscopy}
\label{subsec:pe}
\subsubsection{Theory}\label{subsubsec:petheory}
During PE signal generation, the first two laser–polariton interactions induce either a one-polariton coherence or a population dynamics. The time-dependent polarization evolves over a parametric time interval $t_2$, during which it may undergo transport. The third laser–polariton interaction generates one-polariton coherence components. The time-dependent polarizations generated during intervals $t_1$ and $t_3$ oscillate with opposite phases. When the polariton energy gaps and dephasing parameters are comparable in magnitude, these exponential factors can undergo cancellation, generating an echo. In the presence of polariton transport, an echo additionally requires that the transport matrix elements during the $t_2$ interval be negligible.
The pathways denoted $\text{a1}–\text{a3}$, $\text{a1}'$, and $\text{a2}'$ (Fig.~\ref{fig:vibpolfeyn}) rationalize the physical processes involved. Pathways $\text{a1}$ and $\text{a2}$ involves polariton excited-state absorption and emission, whereas $\text{a3}$ involves polariton ground-state bleaching. The former two are influenced by polariton transport, as illustrated in $\text{a1}'$ and $\text{a2}'$.
Following the derivations outlined in Appendices~\ref{app:response0} and \ref{app:response}, the PE signal is expressed as:
\begin{widetext}
\begin{align}
&S_{\mathrm{PE}}(\Omega_1,\Omega_3) =i^3  c_0 \times  \Big\{ \frac{
d_{p_{2}' p_{0}}^*\mathcal{E}_s^{*}(\omega_{p_{2}' p_{0}}-\omega_s) 
d_{p_{1}' p_{0}}\mathcal{E}_3^{}(\omega_{p_{1}' p_{0}}-\omega_3)  d_{p_{2}' p_{0}}\mathcal{E}_2(\omega_{p_{2}' p_{0}}-\omega_2) 
d_{p_{1}' p_{0}}^*\mathcal{E}_1^*(\omega_{p_{1}' p_{0}}-\omega_1)} 
{(\Omega_3-z_{p_{2}' p_{0}}) (\Omega_1- z_{p_{0} p_{1}'})} \nn\\& 
\times G_{p_{4}'p_{3}'p_{2}'p_{1}'}^{(N)}(t_{2})+
\frac{
d_{p_{2}' p_{0}}^*\mathcal{E}_s^{*}(\omega_{p_{2}' p_{0}}-\omega_s) 
d_{p_{2}' p_{0}}\mathcal{E}_3^{}(\omega_{p_{2}' p_{0}}-\omega_3)  d_{p_{1}' p_{0}}\mathcal{E}_2(\omega_{p_{1}' p_{0}}-\omega_2) 
d_{p_{1}' p_{0}}^*\mathcal{E}_1^*(\omega_{p_{1}' p_{0}}-\omega_1)} 
{(\Omega_3-z_{p_{2}' p_{0}}) (\Omega_1- z_{p_{0} p_{1}'})} G_{p_{0}p_{0}p_{0}p_{0}}^{(N)}(t_{2}) \nn\\&
\frac{
d_{p_{1}''p_{1}'}^*\mathcal{E}_s^{*}(\omega_{p_{1}''p_{1}'}-\omega_s) 
d_{p_{1}'' p_{2}'}\mathcal{E}_3^{}(\omega_{p_{1}'' p_{2}'}-\omega_3)  d_{p_{2}' p_{1}'}\mathcal{E}_2(\omega_{p_{2}' p_{1}'}-\omega_2) 
d_{p_{1}' p_{0}}^*\mathcal{E}_1^*(\omega_{p_{1}' p_{0}}-\omega_1)} 
{(\Omega_3-z_{p_{1}'' p_{1}'}) (\Omega_1- z_{p_{0} p_{1}'})}
G_{p_{4}'p_{3}'p_{2}'p_{1}'}^{(N)}(t_{2})
\Big\}
\label{eqn:pe}
\end{align}
\end{widetext}
The Green's functions $G_{p_{4}'p_{3}'p_{2}'p_{1}'}^{(N)}(t)$, defined in Eq.~(\ref{eqn:gf}), split into two terms describing inter-manifold transport and dephasing contributions. As a result, the signal can be recast as a sum of five distinct pathways. 
The two-dimensional PE spectra are displayed using two scanning frequencies, $\Omega_1$ and $\Omega_3$, for fixed parametric values of the delay time $t_2$. The associated peaks correspond to two distinct types of one-photon transitions: $\omega_{p_j'p_0}$, between the ground and one-polariton manifolds, and $\omega_{p_k''p_j'}$, between the one- and two-polariton manifolds. The $t_2$ dependence is explicitly encoded within the polariton transport matrix elements. 
\subsubsection{Simulations}\label{subsubsec:pesim}
The numerical simulations are presented in Fig.~\ref{fig:vibpolpe} and labeled (A)–(F). For the reference case (A), the central frequencies of the incoming and heterodyning laser pulses follow the energy matching conditions $\omega_1 = \omega_{p_2'p_0}$, $\omega_2 = \omega_{p_2'p_0}$, $\omega_3 = \omega_{p_5'p_0}$, and $\omega_4 = \omega_{p_5'p_0}$. The temporal widths of these pulses set at $\tau_1 =\tau_2 = \tau_3 = 10~\text{fs}$ and $\tau_4 = 10~\text{fs}$. The delay parameter was $t_2= 10~\text{fs}$. 
Three prominent resonance regions are visible along $\tilde{\Omega}_3$, denoted $ (\tilde{\Omega}_{3,a},\tilde{\Omega}_{3,b},\tilde{\Omega}_{3,c})
\approx (19.8, 20.1, 20.7)$; these are further resolved into three sub-regions along $\tilde{\Omega}_1$, denoted $ (\tilde{\Omega}_{1,a},\tilde{\Omega}_{1,b},\tilde{\Omega}_{1,c})
\approx (19.7, 20.2, 20.6)$. The off-diagonal resonances are signatures of polariton energy transfer, while the diagonal peaks indicate polariton mode anharmonicities. The vanishing of off-diagonal peaks suggests the absence of energy transfer in that parameter regime; the distortion of diagonal peaks indicates that polariton excited-state absorption pathways are dominant. Among these resonances, the off-diagonal ones carry the majority of the spectral weights. We use this parameter regime as a reference, and explore several variations below.\\
In case(B), we investigated polariton transport by increasing the delay time to $t_2 = 150$ fs. 
The resulting spectra show that the diagonal peaks at $(\tilde{\Omega}_{1,a},\tilde{\Omega}_{3,a})$ and $(\tilde{\Omega}_{1,c},\tilde{\Omega}_{3,c})$ vanish. Simultaneously, the off-diagonal peak at $(\tilde{\Omega}_{1,c},\tilde{\Omega}_{3,a})$ disappears, while the intensities of the off-diagonal peaks at $(\tilde{\Omega}_{1,b},\tilde{\Omega}_{3,a})$ and $(\tilde{\Omega}_{1,c},\tilde{\Omega}_{3,b})$ are reduced. In this regime, polariton ground-state bleaching and excited-state emission dominate the dynamics. The redistribution of spectral weight due to transport is reflected in the disappearance of the off-diagonal peaks. The resonance-to-state assignments can be  performed via the excitation frequencies specified earlier.\\
In (C), we investigate the role of excitation pulse pairs by setting the central frequencies resonant with the one-polariton state $p'_{3}$, keeping all other parameters the same delay $t_2$ as in (A). The resonances are qualitatively similar to those in (A), with three specific exceptions: the diagonal peak at $(\tilde{\Omega}_{1,a},\tilde{\Omega}_{1,a})$ narrows, the peak at $(\tilde{\Omega}_{3,a},\tilde{\Omega}_{3,a})$ broadens, and the off-diagonal intensities at $(\tilde{\Omega}_{1,a},\tilde{\Omega}_{3,a})$ and $(\tilde{\Omega}_{1,a},\tilde{\Omega}_{3,c})$ diminish. These changes indicate that low-energy polariton transfer is less dominant and that excited-state absorption has shifted to a higher energy domain. These effects would be more pronounced in cases where the two one-polariton states are energetically well-separated.\\
In case (D), we repeated the previous simulation with the delay time extended to $t_2 = 150$ fs. Analogous to the shift from (A) to (B), this case explores the role of transport. The spectra exhibit a qualitative progression similar to the previous pair, though the resonances at $(\tilde{\Omega}_{1,b},\tilde{\Omega}_{3,c})$ and $(\tilde{\Omega}_{1,c},\tilde{\Omega}_{3,b})$ are more pronounced.
It can be inferred that the longer transport timescales lead to a uniform loss of resonances without the generation of new features.
In (E), we explore the effect of narrowband pulses by changing the temporal width of the excitation pulses to $\tau_0 = 40$ fs, while keeping all other parameters identical to those in (A). Although the spectra contain the same features observed in (A), the distribution of spectral weight differs significantly. The peaks at $(\tilde{\Omega}_{1,a},\tilde{\Omega}_{3,b})$ and $(\tilde{\Omega}_{1,a},\tilde{\Omega}_{3,c})$ are the most affected. The peaks at $(\tilde{\Omega}_{1,a},\tilde{\Omega}_{3,b})$ and $(\tilde{\Omega}_{1,a},\tilde{\Omega}_{3,c})$ are the most affected. These narrowband pulses increase the selectivity of the excited-state emission and absorption pathways. Each of these processes involve the two-polariton manifold containing six states. \\
Finally, in (F), we investigate the transport dependence of the narrowband excitation by setting the delay time to $t_2 = 150$ fs. The variation is in the same spirit as those between the pairs (A)–(B) and (C)–(D). The disappearance of the diagonal peaks remains a consistent feature. The spectral weight is redistributed to $(\tilde{\Omega}_{1,a},\tilde{\Omega}_{3,b})$ and $(\tilde{\Omega}_{1,b},\tilde{\Omega}_{3,c})$. In contrast to the previous cases, the off-diagonal peaks are narrower.
\section{Conclusion}
\label{sec:conclusion}
In this work, we investigated the coherent ultrafast dynamics of interacting vibrational modes in the presence of two-vibron nonlinearities, cavity dressing, and vibron-phonon interactions. We deployed two complementary nonlinear spectroscopies that can probe signatures of polariton mode correlations (section~\ref{subsec:dqc}) and polariton transport (section~\ref{subsec:pe}). For each case, we presented two-dimensional frequency correlation plots for six variations of the laser pulse parameters. As a general characteristic, the spectra demonstrate that specific polariton resonances can be amplified or suppressed. It was possible to identify certain parameter regime where dephasing dominates and the shift in resonance is minimal.\\
In section~\ref{subsec:dqc}, we note that the difference between two pathway amplitudes during the time interval $t_{3}$ serves as a discrimination step for the DQC signal. Their relative importance compared with the first pair of coincident pulses is explored in section~\ref{subsubsec:dqcsim}.
In cases where the one-polariton manifold involves multiple states, suitable pulse shaping, such as linearly chirped pulses, can be used to tailor the spectra. The DQC signal probes the nature of two-particle correlations in cavity-dressed vibrons; it can be used to generate a state-resolved map targeting individual two-polariton states \cite{quiros2024resolving, debnath2022entangled, fumero2025biexciton}. 
Further, in section~\ref{subsec:pe}, we observe that narrow resonances survive the effects of transport in most cases. When vibron-phonon couplings are strong or the cavity dressing of these couplings is drastic, this behavior is expected to change. While varying the transport delay, we also note that the diagonal features vanish across the simulation. This uniform feature confirms that the significance of one specific pathway, excited state absorption, is diminished. Shaped laser pulses, such as frequency-chirped pulses (where high-frequency components arrive earlier than low-frequency components within the same pulse profile or vice-versa, depending on the sign of the chirp), offer an opportunity to control the relative amplitudes of ground-state bleaching, stimulated emission, and excited-state absorption pathways.\\
In the final expressions for both the DQC and PE signals (sections \ref{subsubsec:dqctheory} and \ref{subsubsec:petheory}), the effect of cavity enters through the three quantities: transition dipole matrix elements, transition energies, and relaxation parameters. All of them are affected by the exact-diagonalization step, which takes into account the hopping, two-vibron nonlinearity, and cavity-vibron coupling parameters nonperturbatively.
In this work, we interpreted the simulated spectra in the vibron-polariton eigenbasis. Each eigenstate can be expanded as a weighted linear combination of the cavity-vibron tensor-product states. This eigenbasis is convenient for interpreting spectroscopic observables, whereas tensor-product configurations are more intuitive for developing cavity engineering strategies. \\
The tensor-product basis accommodates all configurations relevant to the chosen perturbative order of laser-polariton interactions: zero-quantum excitations, as well as one- and two-quantum excitations residing in vibron, cavity, or joint cavity-vibron modes. For higher-order spectroscopies, expanding the basis set will be necessary. Computational complexity will also increase if a multimode cavity or cubic vibrational nonlinearity (see Eq. \ref{eqn:vibhamabinit}) is included in the initial Hamiltonian.\\
Section~\ref{subsec:vibpoltensor} discusses the cavity-vibron configurations in detail. It establishes a picture how the cavity mediated transitions couples interacting vibrons, generating new basis configurations that are can be probed. In the absence of cavity coupling $g_{mk_c}^{(c)}$, excitation transfer between two vibron modes are governed by the hopping term $J_{mn}$. When a cavity mode is present, energy transfer is governed jointly by $J_{mn}$ and $g_{mk_c}^{(c)}$.
In this work, we made a specific choice about the cavity-vibron coupling strengths, setting them comparable to the energy scales of vibron hopping. Previous investigations have focused primarily on the strong-coupling regime, where coupling strengths match vibron energy gaps, leaving this intermediate regime largely unexplored. The cavity interactions further modify the excited state delocalization induced by the vibron hopping terms, which in turn affects polariton transport and cooperative energy transfer.\\
In multimode vibrons, non-statistical nature of the intermolecular vibrational energy redistribution (IVR) is well-documented \cite{gruebele2004vibrational, keshavamurthy2007dynamical, mondal2024cavity, nesbitt1996vibrational}. The parameter regime examined in this work indicates that tailored cavity modes may enable control of the ultrafast IVR and the corresponding spectroscopic signatures can be monitored. 
The quasiparticle framework will provide a consistent description of these phenomena. However, a full microscopic description requires accurate estimation of the coefficients in Eq.~\ref{eqn:vibhamabinit}. To address these problems, this theoretical framework must be integrated with ab initio simulation methodologies \cite{sidler2024unraveling, li2022qm, flick2018cavity}.\\
One of the principal aims of this work was to clarify the role of higher-order excitations at the ultrafast timescale in the presence of dissipation. The Brownian oscillator model remains valid for more general cases and at finite temperatures. Even though we described the microscopic mechanisms through which cavity coupling influences interacting vibrational dynamics, the detailed description of multilevel vibrations coupled to well-characterized phonon modes is lacking. They are crucial since it may help answer whether excited vibrational modes retain their functional roles as ultrafast spectroscopic markers even in the presence of a cavity \cite{debnath2013dynamics, borges2025selective}.\\
Developing a realistic theory should also account for spatially extended cavity modes, cavity decay and various sources of disorder \cite{engelhardt2023polariton, cao2022generalized, debnath2023theory, lindoy2023quantum, cohn2022vibrational}. 
The cavity decay, for example, can be incorporated directly into the kinetic equations established in appendix \ref{app:gme}. Within the Lindblad formalism, we may add a decay kernel written in the polariton basis $K_{\text{decay}} = \sum_{s} \gamma_s (V_{s}\sigma V_{s}^{\dag}-\frac{1}{2}\{V_{s}^{\dag}V_{s},\sigma \}_+)$, where the manifold-dependent decay rates, $\gamma_s$ are determined by transforming the bare cavity decay rates into the polariton basis. The second anticommutator term in the decay kernel generates the effective Liouvillian, while the first term describes jumps between manifolds. 
The finite lifetime of the cavity modes require a redefinition of the Liouville-space processes, Green's functions. Work in this direction is underway. The theoretical framework can also be extended to explore quantum nonlinear optical measurements in vibrational systems \cite{moradi2025photon}. Research in this direction is currently underway and will be detailed in a forthcoming publication. 
\appendix
\section{Multimode vibron Hamiltonian}
\label{app:ham}
This section supplements the presentation of the Hamiltonian provided in Section~\ref{subsec:vibpolham}. Previous studies developed quasiparticle-based descriptions of localized interacting vibrational modes \cite{mukamel1990femtosecond, christiansen2004vibrational, hamm2011concepts}. The latter provides a rigorous yet tractable description of the underlying physical processes contributing to the time-dependent nonlinear polarization in the weak-driving regime. We begin with the first-quantized Hamiltonian of quantum nuclear dynamics within the Born–Oppenheimer approximation, expressed as
\begin{align}\label{eqn:vibhamabinit}
    H &= \frac{1}{2}\sum_{m}\frac{P_{m}^{2}}{M_{m}} + V(\bm{X})
\end{align}
where $V(\bm{X}) = V(X_1, X_2 , \cdots)$ represents the multidimensional potential energy. The expansion coefficients are related to the derivatives of the potential energy surface (PES) at the equilibrium geometry, which correspond to effective force constants.  To describe the anharmonic dynamics, the potential energy function needs to be expanded around the equilibrium geometry $V_0$, yielding
\begin{align}
& H=\frac{1}{2}\sum_{m} \left(\frac{P_{m}^{2}}{M_{m}}+M_{m}\omega_{m}^{2}X_{m}^{2}\right)+ V_0 +\sum_{m}^{} V_{m}X_{m}\nn\\&
+\sum_{m\ne n}^{}\frac{V_{mn}}{2!}X_{m}X_{n}+\sum_{mnk=1}^{N}\frac{V_{mnk}}{3!}X_{m}X_{n}X_{k}
\nn\\&
+\sum_{mmkl}^{}\frac{V_{mnkl}}{4!}X_{m}X_{n}X_{k}X_{l}+\cdots
\end{align} 
The variables $X (P)$ represent displacement (momentum) coordinates, defined as
\begin{align}
X_{m}=(1/\sqrt{2M_{m}\omega_{m} })(B_{m}^{\dag}+B_{m})
\nn\\
P_{m}=-i\sqrt{M_{m}\omega_{m}/2}(B_{m}^{\dag}-B_{m})
\end{align}
where $B_m^\dag (B_m)$, the creation (annihilation) operators, follow the Bosonic commutation relations.
The Hamiltonian can be recast as,
\begin{align}
H_{\text{vibron},0} &=\sum_{mn}^{} U_{mn}^{}B_{m}^{\dag}B_{n}+\sum_{mnkl}^{}U_{mnkl}^{}B_{m}^{\dag}B_{n}^{\dag}B_{k}B_{l} \nn\\&
+ H_{\text{vibron}, 1}
\end{align}
where 
\begin{align}
& H_{\text{vibron},1}^{} =\sum_{mn}^{} U_{mn}^{} B_{m}^{\dag}B_{n}^{\dag}+ U_{mn}B_{m}B_{n}\nn\\&
+\sum_{mnk}^{}\Big(  U_{mnk} B_{m}^{\dag} B_{n}^{\dag}B_{k}^{\dag}
+U_{mn,k}^{} B_{m}^{\dag}B_{n}^{\dag}B_{k}
\nn\\&
+U_{mnk}^{} B_{m}^{\dag}B_{n}B_{k}+U_{mnk}B_{m}B_{n}B_{k} \Big)
\nn\\&
+\sum_{mnkl} \Big( U_{mnk,l}^{} B_{m}^{\dag} B_{n}^{\dag} B_{k}^{\dag} B_{l}+U_{m,nkl}^{}B_{m}^{\dag}B_{n}B_{k}B_{l}
\nn\\
&+U_{mnkl}B_{m}^{\dag}B_{n}^{\dag}B_{k}^{\dag}B_{l}^{\dag}
+U_{mnkl}B_{m}B_{n}B_{k}B_{l} \Big)
\end{align}
We retained the expansion terms up to the quartic order in the above. The corresponding coefficients are defined as
\begin{align}
&U_{mn}^{}=\delta_{mn}(\omega_{n}+\frac{1}{2}\sum_{l}^{} U_{nnll}^{}) \nn\\
&+\frac{1}{2}(1-\delta_{mn})(\frac{V_{mn}}{\sqrt{(M_{m}\omega_{m})(M_{n}\omega_{n})}}+\sum_{l=1}^{N}U_{mlnl}^{}),
\end{align}
and
\begin{align}
U_{mnkl}^{} &=\frac{1}{2!4!4}\sum_{\text{perm}(mnlk)}\nn\\&
\frac{V_{mnlk}}{\sqrt{(M_{m}\omega_{m})(M_{n}\omega_{n})(M_{1}\omega_{1})(M_{k}\omega_{k})}}
\end{align}
Neglecting number non-conserving terms in $H_{\text{vibron},1}$ and restricting the nonlinearities to the $U_{mm}^{(1)}$ and $U_{mn}^{(2)}$, yields the effective Hamiltonian Eqn.~\ref{eqn:ham} in the main text. The neglected terms represent processes that couple the one- and two-vibron blocks.
\begin{figure*}[ht!]
\begin{center}
\includegraphics[width=.96\textwidth]{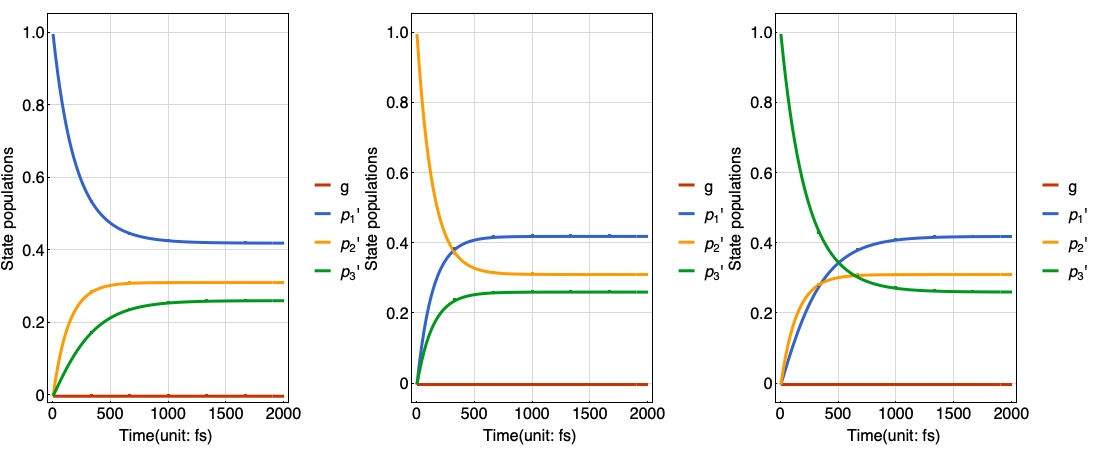}
\end{center}
\caption{Markovian population dynamics of the one-polariton manifold, illustrating the population evolution following initial excitation of each of the three states. Since the system is undriven, the polariton manifolds remain decoupled. The population redistribution reaches a steady state, with the onset of saturation observed around $t = 1.5$~ps.} 
\label{fig:vibpolpopone}
\end{figure*}
\section{Dissipative vibron-polariton kinetics}
\label{app:gme}
\begin{figure*}[ht]
\begin{center}
\includegraphics[width=.96\textwidth]{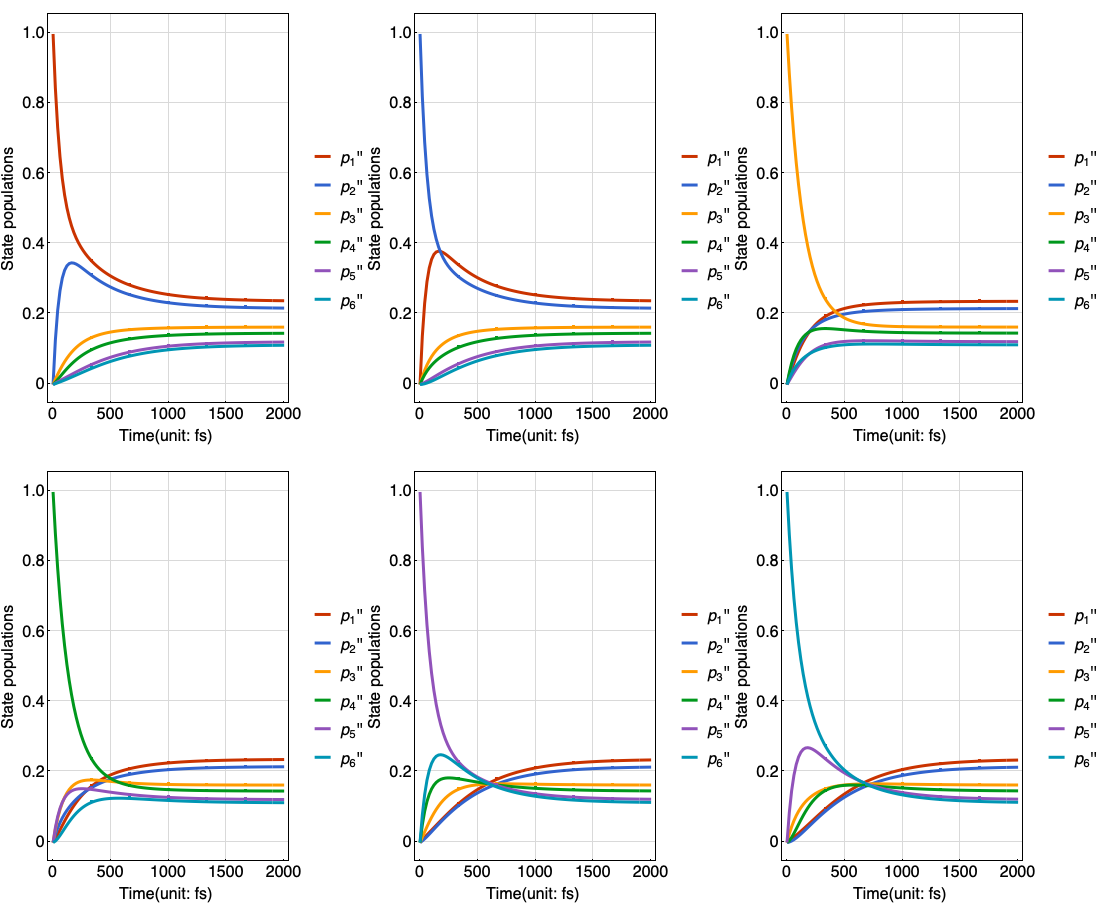}
\end{center}
\caption{Markovian population dynamics of the two-polariton manifold. The panels illustrate population evolution following the initial excitation of each of the six states. Population transfer to the one-polariton manifold is prohibited due to the absence of coupling between two manifolds. Notably, the population redistribution beyond $t=1.5~\text{ps}$ shows no appreciable change. A marked variation in the characteristic decay timescales for each initial state is clearly observed.} 
\label{fig:vibpolpoptwo}
\end{figure*}

This section supplements the description of the Green's functions presented in section~\ref{subsec:vibpolgf}.
Three distinct relaxation parameters, representing polariton transport, dephasing, and pure dephasing, appear in the expressions for the signal. They can be determined starting from the Markovian secular limit of Redfield master equation. Expressed in the polariton eigenbasis, it reads
\begin{align}\label{eq:gme}
    \frac{d}{dt} \,\sigma_{ab} &= -i [H_{S},\sigma]_{ab}+\sum_{cd}K_{ab,cd} \, \sigma_{cd}
\end{align}
The tetradic relaxation kernel, $K_{ab,cd}$, accounts for all relaxation effects originating from polariton-phonon interactions. It is defined as:
\begin{align}\label{eqn:redfield}
&K_{a_{4}a_{3},a_{2}a_{1}}=\sum_{j}\lambda_{j}\{\delta_{a_{1}a_{3}}\sum_{a^{\prime}}\overline{M}_{j}^{(+)}(\omega_{a_{2}a^{\prime}})  \nn\\
&\times\Phi_{a_{4}a^{\prime}a_{2}a^{\prime}}
+\delta_{a_{4}a_{2}}\sum_{a^{\prime}}\overline{M}_{j}^{(-)}(\omega_{a^{\prime}a_{1}})\Phi_{a^{\prime}a_{3}a^{\prime}a_{1}}-\nn\\
&\overline{M}_{j}^{(+)}(\omega_{a_{2}a_{4}})\Phi_{a_{4}a_{3}a_{2}a_{1}}-\overline{M}_{j}^{(-)}(\omega_{a_{3}a_{1}})\Phi_{a_{4}a_{3}a_{2}a_{1}}\}
\end{align}
Each term in the expression is weighted by the overlap matrix elements
\begin{align}
    \Phi_{a_4 a_3 a_1 a_1} &= \sum_n T_{na_4} T_{na_3} T_{na_2} T_{na_1}
\end{align}
and the relaxation function, defined as the half-Fourier transform of the bath correlation function,
\begin{align}
\overline{M}_{j}^{(\pm)}(\omega_{a_1 a_2})=\int_{0}^{\infty}d t \exp(i\omega_{a_1 a_2} t)\, \overline{C}_{j}^{(\pm)}(t)
\end{align}
The expression for $\overline{M}_{j}^{(\pm)}(\omega_{a_1 a_2})$ represents the phonon spectral function sampled at the polariton energy gaps. As a consequence, only the resonant polariton-phonon interactions govern the dissipation.
The phonon spectral function, specified in Eq.~\ref{eqn:phononspecf} is related to the correlation function as follows
\begin{align}
\overline{C}_{}^{(\pm)}(t)=&\int_{-\infty}^{\infty}\frac{d\omega}{2\pi}J_{}(\omega) N_\beta(\omega)\times\nn\\
&[\coth(\beta\hbar\omega/2)\cos(\omega t)\mp i~\sin(\omega t)]
\end{align}
By evaluating these integrals, we obtain the time-domain correlation function 
\begin{align} 
   & C_{}(\tau)= (\lambda_0 \gamma_0/2) \cot(\beta \gamma_0/2)\exp(-\gamma_0 \tau)+
   \frac{\lambda_{}}{2\zeta_{}} \nonumber\\
   &\Big(\coth(i \beta \phi_{}^{+}/2)\exp(-\phi_{}^{+} \tau)-\coth(i \beta \phi_{}^{-}/2)\exp(-\phi_{}^{-} \tau)\Big)\nonumber\\
   &+(-i\lambda_0 \gamma_0/2) \exp(-\gamma_0 \tau)+ \frac{i\lambda_{} \upsilon_{}^2}{2\zeta_{}}\Big(\exp{(-\phi_{}^{+}t)}-\nonumber\\
   &\exp{(-\phi_{}^{-}t)}\Big)
 -\sum_{n=1}^\infty\Big((4\lambda_{}\gamma_{} \upsilon_{}^2/\beta) (\nu_n/(\upsilon_{}^2+\nu_n^2)^2-\nu_n^2\gamma_{}^2) \nonumber\\
   &+(2\lambda_0\gamma_0/\beta) (\nu_n/(\nu_n^2-\gamma_0^2)) \Big)
   \exp(-i \nu_n \tau)
\end{align}
The parameters are defined as $\phi_{}^{\pm} =(\gamma_{}/2)\pm i \zeta_{}$, where $\zeta_{}=\sqrt{ (\upsilon_{}^2-\gamma_{}^2/4)}$. The thermal environment is characterized by the inverse temperature $\beta=1/\kappa T$, where $\kappa$ and $T$ denote the Boltzmann constant and the temperature, respectively. The Matsubara frequencies, $\nu_{n}=n (2\pi/\beta)$ result from the partial fraction expansion of the $\coth(\beta\hbar\omega/2)$ term in the spectral density. This formulation rigorously incorporates finite-temperature effects.\\ 
From the kinetic equation specified above, we can obtain the polariton transport kernel
\begin{align}\label{eqn:paramtransport}
K_{a_2 a_2,a_1 a_1} &=-2\mathrm{Re}\sum_{j}\lambda_{j}\overline{M}_{j}^{(+)}(\omega_{a_1 a_2})\Phi_{a_2 a_2 a_1 a_1},
\end{align}
This kernel governs the time evolution of the population terms of the polariton density operator via
\begin{align}\label{eqn:eomtransport}
\frac{d}{dt}\rho_{a_2 a_2}(t) &=-\sum_{a_1}K_{a_2 a_2,a_1 a_1}\rho_{a_1 a_1}(t),
\end{align}
The matrix elements of the kernel obey probability conservation, $\sum_{a_1}K_{a_1 a_1,a_2a_2}=0$, and the detailed balance condition
\begin{align}
   K_{a_2a_2,a_1a_1}/K_{a_1a_1,a_2a_2} &=\exp(-\omega_{a_2a_1}/(k_{B}T)) 
\end{align}
The transport Green's function is given by
\begin{align}\label{eqn:gftransport}
G_{a_2 a_2,a_1 a_1}^{(N)}(t) &=[\exp(-K t)]_{a_2 a_2,a_1a_1}\nn\\
&\equiv\sum_{r}\chi_{a_2 r}^{(R)}D_{rr}^{-1}\exp(-\lambda_{r}t)\chi_{r a_1}^{(L)},
\end{align}
where $\lambda_r$ is the $r$-th eigenvalue of the transport kernel $K_{a_2 a_2,a_1a_1}$. The matrices $\chi^R (\chi^L) $ denote the right (left) eigenvectors, which satisfy the relation $D_{}=\chi^L\chi^R$ and normalization condition $\chi^L\chi^R=1$. The transport kernel affects the photon-echo signal.
The transport kernel can be used to track the time evolution of the population terms. Using initial state vectors that represent excitations to each one- and two-polariton state, we present the results in Fig.~\ref{fig:vibpolpopone} and Fig.~\ref{fig:vibpolpoptwo}, respectively. These figures provide early intuition regarding the nature of population evolution at various waiting times.\\
From the kinetic equation, we can also obtain the polariton coherence kernel. It satisfies the following equation of motion
\begin{align}\label{eqn:eomdephasing}
\frac{d}{dt}\rho_{a_1a_2}(t)=[-i\omega_{a_1 a_2}-\gamma_{a_1 a_2}^{(N)}]\rho_{a_1 a_2}(t), a_1\ne a_2,
\end{align}
The dephasing parameter is defined as
\begin{align}\label{eqn:paramdephasing}
\gamma_{a_1a_2}^{(N)}=\frac{1}{2}(K_{a_1a_1,a_1a_1}+K_{a_2a_2,a_2a_2})+\tilde{\gamma}_{a_1a_2}.
\end{align}
where the state-specific line-broadening function is given by
\begin{align}\label{eqn:parampuredephasing}
\gamma_{a_1} &=\sum_{j}\lambda_{j}\sum_{a_2}\overline{M}_{j}^{(+)}(\omega_{a_1 a_2})\Phi_{a_1a_1 a_2 a_2}
\end{align}
The corresponding coherence Green's function is
\begin{align}\label{eqn:gfdephasing}
    G_{a_1a_2}(t) &= \exp{[-i\omega_{a_1a_2} t-\gamma_{a_1a_2}^{(N)} t]}
\end{align}
It affects the photon echo spectroscopy. A compact form combining the population and coherence Green's functions is presented in Eq.~\ref{eqn:gf}.\\
Due to the number conservation of the Hamiltonian in the absence of the laser driving, the relaxation kernel in Eq.~\ref{eqn:redfield} can be factorized for each manifold. The quantities derived in Eq.~\ref{eqn:paramtransport}, \ref{eqn:paramdephasing}, and \ref{eqn:parampuredephasing} represent intra-manifold relaxation parameters. In contrast, the inter-manifold dephasing parameters are estimated using individual line-broadening functions
$\gamma_{a_1' a_2^{}}^{} =(\gamma_{a_1'}+\gamma_{a_2^{}})/2$. 
The corresponding Green's function is given by
\begin{align}\label{eqn:gfdephasinginter}
    G_{a_1'a_2,a_1'a_2}(t) &= \exp{[-i\omega_{a_1'a_2} t-\gamma_{a_1' a_2^{}}^{} t]}
\end{align}
where the primed and unprimed indices denote different manifolds. This expression is relevant for both photon echo and double-quantum coherence spectroscopy.
\section{Nonlinear response formalism for finite pulses}
\label{app:response0}
This section introduces the nonlinear response formalism and complements section~\ref{sec:mdcs}. The laser-induced nonlinear polarization is defined as the expectation value of the dipole operator $d$: 
\[P^{(n)}= \mathrm{Tr}[d_{}\rho^{(n)}]\] where $\rho^{(n)}$ represents the $n$-th order laser-perturbed polariton density operator. In the perturbative regime, the density operator is expanded in orders of the laser–polariton interaction to yield 
\begin{align}
    &\rho(t) = G(t)\rho(0)+\sum_{n=1}^{\infty}(-i)^{n} \int_{0}^{t}d\tau_{n} \cdots\int_{0}^{\tau_{2}}d\tau_{1}\times \nn\\&
G(t-\tau_{n})L_{\text{int},-}^{}(\tau_{n}) \cdots
G(\tau_{2}-\tau_{1})L_{\text{int}}^{}(\tau_{1})G(\tau_{1})\rho(0)
\end{align}
dipole superoperator is defined as $L_{\text{int},-}^{} O = [H_{\text{int}}^{}, O] $. The PE and DQC signals are related to the third-order ($n=3$) nonlinear polarization induced by three incoming laser pulses
\begin{align}
    E(t) &=\sum_{j, u_{j}=\pm 1}\mathcal{E}_{j}^{u_{j}}(t-\tau_{j})\exp[i u_{j}(\bm{k}_{j} \bm{r}-\omega_{j}(t-\tau_{j}))]+ \text{c.c.}
\end{align}
The field profile of the $j$-th pulse is given by $\mathcal{E}_{j}^{u_{j}}(t - \tau_{j})$. Here, $\omega_j$ denotes the carrier frequency, $\mathbf{k}_j$ represents the wavevector, and $\tau_j$ is the centering time. The index $u_{j}$ tracks the distinct frequency and wavevector components of the incoming field.\\ 
The third-order nonlinear polarization is expressed as
\begin{align}
    P^{(3)}(\bm{r},t)=\sum_{\bm{k}_{s}}P_{\bm{k}_{s}}^{(3)}(t)\exp[i \bm{k}_{s} \bm{r}]
\end{align}
where
\begin{align}
     P_{\bm{k}_{s}}^{(3)}(t)
          &= \exp[-i \omega_s T_{3} 
- i (u_2 \omega_2 + u_1 \omega_1) T_{2} - i u_1 \omega_1 T_{1}] \nn\\&
\int_{0}^{\infty}\int_{0}^{\infty}\int_{0}^{\infty}dt_{3}dt_{2}dt_{1}
R_{k_{s}}^{(3)}(t_{3},t_{2},t_{1})\times
\nn\\&  
\exp[i\omega_{s}t_{3}+i(u_{2}\omega_{2}+u_{1}\omega_{1})t_{2}+iu_{1}\omega_{1}t_{1}] \nn\\&
    \mathcal{E}_{3}^{u_{3}}(t-t_{3}-\tau_{3}) 
    \mathcal{E}_{2}^{u_{2}}(t-t_{3}-t_{2}-\tau_{2})\nn\\&
    \mathcal{E}_{1}^{u_{1}}(t-t_{3}-t_{2}-t_{1}-\tau_{1})
\end{align}
The variables $t_j = \tau_{j+1} - \tau_j$ denote the controllable delays between the incoming laser pulses1. The third-order time-domain response function, $R^{(3)}(t_3, t_2, t_1)$, contains the complete information regarding the vibron-polariton dynamics. For PE and DQC spectroscopies, the response functions correspond to the specific pathways depicted in Fig.~\ref{fig:vibpolfeyn}. The PE ($\bm{k}_{\text{I}}$) and DQC ($\bm{k}_{\text{III}}$) signals are selected in the phase-matching direction
$\bm{k}_{\text{I}}=- \bm{k}_{1}+ \bm{k}_{2}+\bm{k}_{3}$ and $\bm{k}_{\text{III}}=+ \bm{k}_{1}+ \bm{k}_{2}-\bm{k}_{1}$, respectively. \\

The third-order signal is typically detected via heterodyne techniques using a local oscillator field $E_{s}^{*}(t-\tau_{s})$. The signal in the time-domain can be expressed as,
\begin{align}
S(T_{3},T_{2},T_{1}) &= \int_{-\infty}^{\infty} dt P_{\bm{k}_{s}}^{(3)}(t)
E_{s}^{*}(t-\tau_{s}),
\end{align}
where $T_3=\tau_s-\tau_3$, $T_2=\tau_3-\tau_2$, and $T_1=\tau_2-\tau_1$ denote the time delay variables. Finally, the two-dimensional spectrogram representation of the signal is obtained by performing a joint Fourier transform with respect to the delay variables. It is expressed as
\begin{align}
& S_{\bm{k}_{s}}^{(3)}(\Omega_{3},\Omega_{2},\Omega_{1}) =
\int_{0}^{\infty}dT_{3}\int_{0}^{\infty}dT_{2}\int_{0}^{\infty}dT_{1}
\nn\\& 
\exp[{i\Omega_{3}T_{3}+i\Omega_{2}T_{2}+i\Omega_{1}T_{1}}] S_{\bm{k}_{s}}^{(3)}(T_{3},T_{2},T_{1})
\end{align}
\section{Nonlinear response functions for the photon-echo and double quantum coherence signal}
\label{app:response}
This section supplements section~\ref{sec:mdcs}, especially the derivation of the signal in section~\ref{subsubsec:dqctheory} and \ref{subsubsec:petheory}. The response functions are expanded in the field-free polariton eigenbasis. While this expansion is numerically demanding for large systems, it allows the initial equilibrium density operator to be expressed consistently. For systems multimode system interacting with multiple cavity modes, this approach may become numerically intractable. In such cases, a variational search procedure is more appropriate for finding the initial state.\\
The time-domain PE response function is expressed as 
\begin{widetext}
   \begin{align}
R_{k_{\text{I}}}^{}(t_{3},t_{2},t_{1}) &=
i^{3}\theta(t_{1})\theta(t_{2})\theta(t_{3})\sum_{p_{4}'p_{3}'p_{2}'p_{1}'}d_{p_{0} p_{4}'}d_{p_{3}'p_{0}}d_{p_{2}'p_{0}}d_{ p_{0}p_{1}'} \exp(-i z_{p_{4}'p_{0}}t_{3}-i z_{p_{0}p_{1}'}t_{1}) G_{p_{4}'p_{3}'p_{2}'p_{1}'}^{(N)}(t_{2}) \nn\\&
+i^{3}\theta(t_{1})\theta(t_{2})\theta(t_{3})\sum_{p_{4}',p_{1}'}d_{p_{0}p_{4}'}d_{p_{4}'p_{0}}d_{p_{1}'p_{0}}d_{p_{0}p_{1}'} \exp(-i z_{p_{4}'p_{0}}t_{3}-i z_{p_{0} p_{1}'}t_{1}) \nn\\&
-i^{3}\theta(t_{1})\theta(t_{2})\theta(t_{3})
\sum_{p_{1}''p_{4}'p_{3}'p_{2}'p_{1}'}d_{p_{3}'p_{0}}d_{p_{1}''p_{4}'}d_{p_{2}'p_{0}}d_{p_{0}p_{1}'}
\exp(-i z_{p_{1}''p_{4}'}t_{3}-i z_{p_{0} p_{1}'}t_{1})G_{p_{4}'p_{3}'p_{2}'p_{1}'}^{(N)}(t_{2})
\label{eqn:petime}
\end{align} 
\end{widetext}
It contains contributions from three terms, containing five pathways. The corresponding diagrams displayed in Fig.~\ref{fig:vibpolfeyn} help rationalize them. The two-dimensional spectra is displayed by taking a Fourier transform w.r.t. $T_1$ and $T_3$. The final expression of PE signal is presented in Eq.~\ref{eqn:pe}.
The time-domain DQC response function is expressed as,
\begin{widetext}
   \begin{align}
R_{k_{\text{III}}}^{}(t_{3},t_{2},t_{1})&= -i^{3}\theta(t_{1})\theta(t_{2})\theta(t_{3})\sum_{}d_{p_{2}''p_{1}''}d_{p_{0}p_{2}'}d_{p_{1}''p_{1}'}d_{p_{1}'p_{0}}\exp(-i z_{p_{1}''p_{2}'}t_{3}-i z_{p_{1}''p_{0}}t_{2}-i z_{p_{1}'p_{0}}t_{1}) \nn\\&
+i^{3}\theta(t_{1})\theta(t_{2})\theta(t_{3})\sum_{}d_{p_{0}p_{2}'}d_{p_{2}'p_{1}''}d_{p_{1}''p_{1}'}d_{p_{1}'p_{0}}\exp(-i z_{p_{2}'p_{0}}t_{3}-i z_{p_{1}''p_{0}}t_{2}-i z_{p_{1}'p_{0}}t_{1})
\label{eqn:dqctime}
\end{align} 
\end{widetext}
It contains contributions from two distinct pathways, which can rationalized from the diagrams in Fig.~\ref{fig:vibpolfeyn} in a similar manner. The two-dimensional spectra is displayed by taking a Fourier transform w.r.t. $T_2$ and $T_3$. The final expression of DQC signal is presented in Eq.~\ref{eqn:dqc}.
\begin{acknowledgments}
A.D. acknowledges support from DESY (Hamburg, Germany), a member of the Helmholtz Association HGF. 
A.D. acknowledges technical discussions with A. Rubio (MPSD, Hamburg) regarding ab initio simulation algorithms of vibron-polariton dynamics.
\end{acknowledgments}

\bibliography{ref_vibron_polariton}

\end{document}